# A congested schedule-based dynamic transit passenger flow estimator using stop count data

**Qi Liu, Joseph Y. J. Chow***

*Department of Civil and Urban Engineering, New York University, Brooklyn, NY, 11201, USA*

*Corresponding author email: joseph.chow@nyu.edu

**ABSTRACT**

A dynamic transit flow estimation model based on congested schedule-based transit equilibrium assignment is proposed using observations from stop count data. A solution algorithm is proposed for the mathematical program with schedule-based transit equilibrium constraints (MPEC) with polynomial computational complexity. The equilibrium constraints corresponding to the schedule-based hyperpath flow are modified from the literature to fit into an estimation problem. Computational experiments are conducted first to verify the methodology with two synthetic data sets (one of which is Sioux Falls), followed by a validation of the method using bus data from Qingpu District in Shanghai, China, with 4 bus lines, 120 segments, 55 bus stops, and 120 one-minute intervals. The estimation model converged to 0.005 tolerance of relative change in 10 iterations. The estimated average of segment flows are only 2.5% off from the average of the observed segment flows; relative errors among segments are 42.5%.

# 1 INTRODUCTION

Transit network information like time-varying origin-destination (OD) demand, passenger flows, and run ridership is indispensable for transit operations. Despite the importance of this information, many transit operators do not have access to data sources like smartcard or cellular phone that can be used to conveniently infer user trajectories hence transit run flows. This is possibly due to budget reason for small transit networks or privacy concerns.

Passenger flow estimation, i.e. of the ridership volumes on each run in the system when APC use is limited, is a complex problem. For one, the problem is generally more complex than other traffic flow estimation problems because of the added scheduling of transit lines, added walking, waiting, transfer activities, and limited data. It is clear there are two distinct classes of passenger estimation models: planning models and real-time operational models. The former use limited count data—turnstile or boarding/alighting counts and AFC data—and seek to determine at least linked transit journeys (e.g. Gordon et al. (2013), Sun et al. (2015), Sun and Schonfeld (2015)). The results are used for strategic planning.

Transit operational prediction models, on the other hand, are applicable to many operational scenarios: station queue monitoring, incident management, holding and rescheduling strategies, and more (see Sun and Schonfeld (2015)). However, transit operational passenger flow estimation is more complex because more sophisticated schedule-based transit assignment models are needed due to the transient nature of the problem. The presence of queueing in dynamic networks makes the passenger flow estimation a nonconvex problem. Zhu et al. (2017) and Zhang et al. (2017) handle estimation at either line or station level, not at network level. While other technologies exist to allow more sophisticated estimation using mobile communications data, e.g. Aguiléra et al. (2014) and de Regt et al. (2017), we focus on what can be estimated using only stop-based count data since many systems do not have access to the mobile device data.

We propose the first schedule-based transit passenger flow estimator under congestion using station count data, which includes estimating OD demand and line flows. The estimation model can be used for evaluating operational strategies by providing dynamic passenger flows. Section 2 elaborates on the research gap for a dynamic passenger flow estimation model. Section 3 covers the methodology, which includes a new dynamic passenger flow estimation model as a mathematical program with schedule-based transit equilibrium constraints (MPEC), along with two solution algorithms. The equilibrium constraints corresponding to the schedule-based hyperpath flow is adapted from the literature with some modifications to fit it into an estimation problem. Properties of the MPEC and its solution algorithms are discussed. Section 4 covers computational experiments to verify the methodology with two synthetic data sets (one of which is Sioux Falls). Section 5 presents a validation of the method using data from Qingpu District in Shanghai, China. Section 6 concludes.

# 2 LITERATURE REVIEW

Early literature on network estimation started with OD estimation without consideration of passenger assignment. Measurements are used as constraints; the goals include maximizing entropy (Macgill, 1977), minimizing information (Willekens, 1980), etc. In cases where measurements can be inconsistent, the goal is to maximize likelihood (Van

Zuylen and Willumsen, 1980) or minimize least squares in errors (Cascetta, 1984). Time series tools have also been applied for OD flow estimation; Cremer and Keller (1987), Nihan and Davis (1987) have sought to use statistical techniques to estimate O/D matrices. Some online OD flow estimation based on Kalman filter have been proposed (Okutani and Stephanedes, 1984, Ashok and Ben-Akiva, 2000). Other methods include network tomography approaches (Zhang et al., 2017) and network-oriented machine learning like multi-agent inverse optimization (Xu et al., 2018). Kumar et al. (2018) used smart card data to infer passenger paths and OD matrices.

User equilibrium (UE) assignment-based estimation models are used to account for traveler behavior. In such models, OD flows are estimated such that their UE assignment best fits the count data. Such models combining estimation error minimization and UE condition have been considered (Yang et al., 1992), but calling it a proper bilevel problem, i.e. Stackelberg game (see Bard (2013)), does not make sense because the upper level problem is not technically a separate decision-maker making changes to the network. It is rather an estimation problem with equilibrium constraints. Indeed, further study by Yang (1995) showed the estimation results using algorithms assuming a Cournot-Nash equilibrium versus those assuming a Stackelberg game are indistinguishable in performance of accuracy. Hence, this is a mathematical program with equilibrium constraints (MPEC) but not a bilevel problem.

As for models of the UE constraints, they are further classified into static UE and dynamic UE. Static models assume that the demand, service rate and congestion level are constant throughout the analysis period. For transit, static UE typically refers to frequency-based models. The frequency-based approach assumes stochastic transit arrival times and that users choose from different common lines to form a preference set. It is typically assumed that users choose the first arriving line in the set, following the optimal strategy principle set by Spiess and Florian (1989). This leads to probabilistic paths, called hyperpaths, across multiple routes when the time dimension is ignored. The existence of asymmetric interactions leads to a variational inequality (VI) problem.

Transit assignment models can also be schedule-based (Wilson and Nuzzolo, 2013, Chin et al., 2016). Schedule-based models assume that users choose from transit runs at transfer nodes. These available runs form an ordered set, sometimes called an ordered preference set; users always choose the most preferable run that is feasible for boarding. Whereas the frequency-based approach is simpler, schedule-based models are inherently dynamic; the evolution of the transit system state over time is tracked. Hamdouch and Lawphongpanich (2008), Hamdouch et al. (2004), Hamdouch et al. (2011), (Nuzzolo et al., 2012) studied the dynamic UE for schedule-based transit assignment. In these models, rigid capacities are imposed and priorities among flows are considered. Some studies use simulation methods instead to evaluate within day flow dynamics. Cats and West (2020) model the within day dynamic path choice by means of day-to-day learning implemented in an agent-based simulation model.

It is apparent that passenger flow estimation for transit passengers needs to be dynamic for operational use. Wong and Tong (1998) proposed a maximum entropy estimator; Nuzzolo and Crisalli (2001) proposed a least squares estimator; they both assume schedule-based transit service with no congestion effect. Postorino et al. (2004) compared OD estimation model under assumptions of frequency-based and schedule-based assignment for uncongested networks. Lam et al. (2003), Lam and Wu (2004), and Wu and Lam (2006) proposed least squares OD estimators with frequency-based equilibrium assignment constraints with congestion effects. Montero et al. (2015) proposed real-time transit OD flow estimation formulated as Kalman filtering; proportions of passenger traversing links are calculated by assigning the most likely OD demand matrix; these proportions are updated when new measurements are available. The above models are all dynamic, meaning that they estimate time-dependent OD flows. There are studies on transit flow estimation using various data sources, like smartcard data (Pelletier et al., 2011, Tao et al., 2014, Yang et al., 2020) and cellular phone data (Sohn and Kim, 2008, Caceres et al., 2012). In this study, we focus on estimation using the most basic type of data - count data. These studies are summarized in Table 1.

As we can see, the literature in transit passenger flow estimation has not considered dynamic OD and flow estimation with congested schedule-based transit equilibrium constraints.

Table 1. Summary of transit OD flow estimation models using count data

| Studies | Uncongested (U) or Congested (C) | Frequency-based (F) or Schedule-based (S) | (Upper-level) goals |
| --- | --- | --- | --- |
| Nguyen et al. (1988) | U | F | Max entropy & max likelihood |
| Wong and Tong (1998) | U | S | Max entropy |
| Nuzzolo and Crisalli (2001) | U | S | Least squares |
| Postorino et al. (2004) | U | F & S | Least squares |
| Lam et al. (2003) | C | F | Least squares |
| Lam and Wu (2004) | C | F | Least squares |
| Wu and Lam (2006) | C | F | Least squares |
| Montero et al. (2015) | - | - | Least MSE (Kaman Filtering) |

## 3 PROPOSED METHODOLOGY

Our proposed model is a least squares estimation model where the underlying schedule-based transit equilibrium constraint is an extension of Nuzzolo and Crisalli (2001) model to congested schedule-based transit networks. Consider a transit network of a set of stops $N$ traversed by a set of lines $L$ over a time horizon $T$ set of intervals. A time-expanded (TE) network (Ahuja et al., 1993) is used for the transit system representation, where each TE node is $i_h \coloneqq (i, h) \in \mathcal{N} \times \mathcal{T}$. A hyperpath is a subgraph of the TE-network composed of an ordered preference set of attractive lines for each current passenger location $i_h$ to their destination. The measurements include entry, exit and pass-by counts at stops (via stop-based IoT sources like Wi-Fi). Count data are aggregated and reported periodically. The counting period of different measurements are assumed to be the same, although this can be trivially extended. The network and model representation are illustrated in Figure 1. It shows one hyperpath to destination node *N5* and how two users departing from *N1* at time 0 and 1 respectively get to destination *N5*. Stop flows are reported every three time units at each stop.

*Notation: parameters*
$\mathcal{N}$: Set of all physical transit stops; $N \coloneqq \text{card}(\mathcal{N})$, the cardinality of $\mathcal{N}$;
$\mathcal{H}$: set of points in time; horizon $H \coloneqq \text{card}(\mathcal{H})$;
$\mathcal{L}$: Set of transit lines; $L \coloneqq \text{card}(\mathcal{L})$;
$\mathcal{K}$: Set of measurements; $K \coloneqq \text{card}(\mathcal{K})$;
$\mathcal{T}$: Set of discrete time slices; $T \coloneqq \text{card}(\mathcal{T})$ is the analysis horizon;

$\mathcal{W}$: Set of tuples $(q, r, h)$ where $q \in N$ means the origin, $r \in N$ means the destination and $h \in T$ is the departure time; $W := \text{card}(\mathcal{W})$;
$\mathcal{S}_w$: Set of hyperpaths for OD tuple $w \in \mathcal{W}$;
$\mathcal{S}$: Set of all hyperpaths, namely $\mathcal{S} = \bigcup_{w \in \mathcal{W}} \mathcal{S}_w$;
$S$: the maximum number of hyperpaths for each $w$;
$d^w$: Demand for $w \in \mathcal{W}$;
$\boldsymbol{d}$: Demand vector; $\dim(\boldsymbol{d}) = W$;
$\mathcal{A}$: Set of links on TE-network; $A := \text{card}(\mathcal{A})$;
$\mathcal{A}^+(i_h)$: set of arcs emanating from node $i_h$;
$t_a$: Travel time of link $a \in \mathcal{A}$;
$c_a$ Travel cost of link $a \in \mathcal{A}$;
$C(\cdot)$: Total cost travel cost of the network, a function of hyperpath flow vector $\boldsymbol{f}$;
$x_{i,h}$: The quantity of entry/exit/pass-by flow at stop $i \in N$ at time $h$, an intermittent variable;
$X_{i,k}$: $k$-th aggregated measurement of entry/exit/pass-by flow at stop $i \in N$;
$\Omega$: Set of all feasible hyperpath flow vectors.

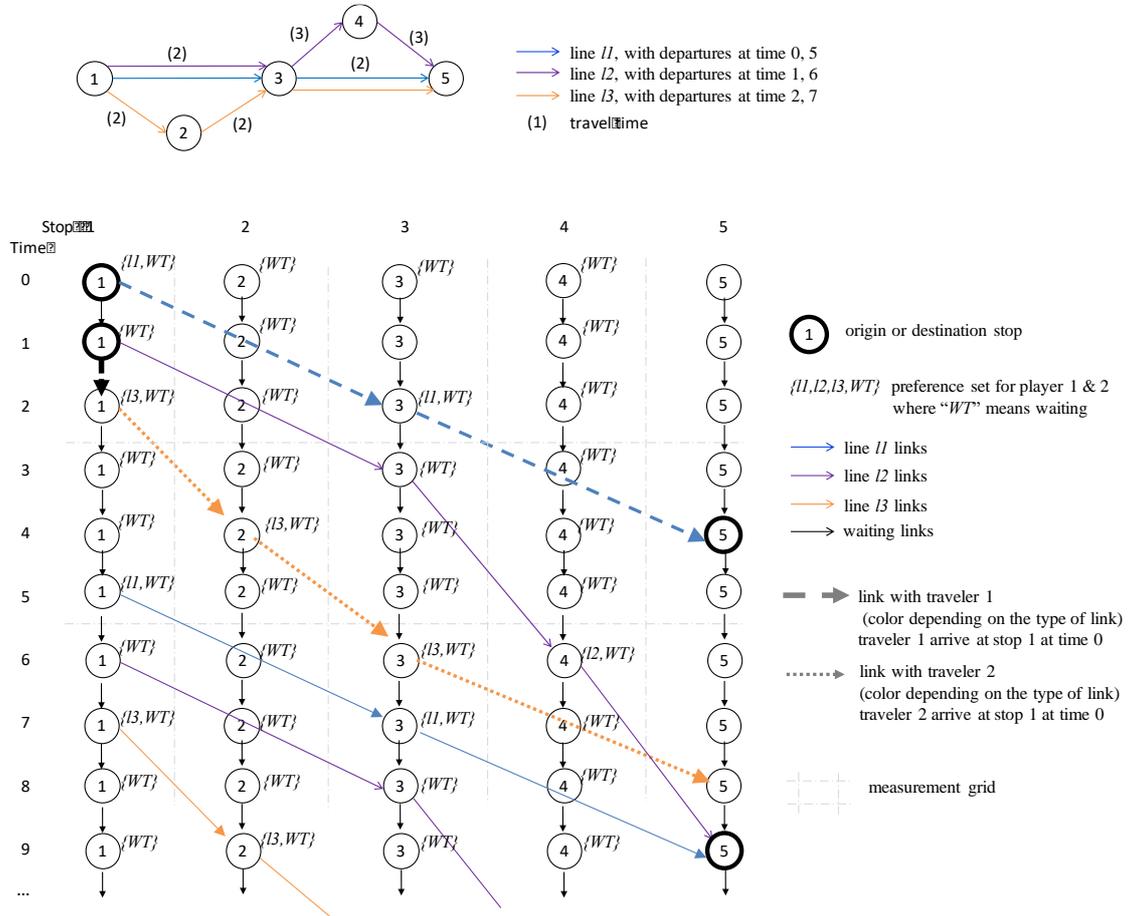

Figure 1. Illustration of time-expanded network and terminology used.

*Notation: decision variables*
$f^s$: Flow for hyperpath $s \in \mathcal{S}_w$;
$\boldsymbol{f}$: hyperpath flow vector; $\dim(\boldsymbol{f}) = \text{card}(\Omega)$;

$p_{i,h}^w$: The proportion of flow from OD $w \in \mathcal{W}$ entering/exiting/passing-by stop $i$ at time $h$;

$z^{s,l,\tau,i_h}$: The fractional hyperpath $s \in S_w$ flow appearing at stop $i$ at time $h$ with the arriving route being $l \in \mathcal{L}$ and arriving time being $\tau \in [0, h]$;

$y^{s,l,\tau,a}$: The fractional hyperpath $s \in S_w$ flow passing link $a$ with the arriving route being $l \in \mathcal{L}$ and arriving time being $\tau \in \mathcal{T}$;

### *3.1 Proposed dynamic UE Assignment-based passenger flow estimator*

A MPEC structure is adopted. The objective function (1a) is to minimize the squared errors of the flows at each stop. The intermittent variable $x_{i,h}$ represents entry, exit or pass-by flow at stop $i$ at time $h$; $X_{i,k}$ is the measurement of the $x_{i,h}$ during the $k$-th period. $x_{i,h}$ is related to OD demand $\boldsymbol{d}$ through the detection probabilities $\boldsymbol{P}$. Ordinary least squares (OLS) is used to estimate the path flows.

Eq. (1b) is used to describe the relationship between demand variables and intermittent variables. $p_{i,h}^w$, an element of $\boldsymbol{P}$, refers to the proportion of flow from OD $w$ measured at stop $i$ at $h$. $\boldsymbol{P}$ is determined by the lower level schedule-based UE model, formulated as variational inequality constraints shown in Eq. (1c) – (1d).

$$\min_{\boldsymbol{d}} \sum_{i \in \mathcal{N}, k \in \mathcal{K}} \left( \sum_{h=h_{k-1}}^{h_k} x_{i,h} - X_{i,k} \right)^2 \tag{1a}$$

$$\sum_{w \in \mathcal{W}} p_{i,h}^w d^w = x_{i,h}, \quad \forall i \in \mathcal{N}, h \in \mathcal{H} \tag{1b}$$

$$\boldsymbol{d} \geq \boldsymbol{0}$$

Equilibrium constraints:

$$C(\boldsymbol{f})(\tilde{\boldsymbol{f}} - \boldsymbol{f})^T \geq 0, \quad \forall \tilde{\boldsymbol{f}} \in \Omega \tag{1c}$$

$$\sum_{s \in S_w} f^s = d^w, \quad \forall w \in \mathcal{W} \tag{1d}$$

$$\boldsymbol{f} \geq \boldsymbol{0}$$

### *3.2 Solution Algorithm for the MPEC*

To solve this MPEC model, we adopt the *link usage proportion-based algorithm* from Yang et al. (1992), Yang and Bell (1998), by separating the problem into two subproblems: the primary estimation subproblem problem in Eq. (1) without equilibrium constraints Eq. (1c) – (1d) and with fixed proportions, and the equilibrium problem with fixed OD demand. A proportion represents the ratio of passengers for a particular OD flow passing through a particular link. The algorithm, which we call the Alternate Updating Method (AUM), is shown below.

ALGORITHM AUM (Alternate Updating Method)
---
*Step 1*: Initialize proportions $P^{(0)}$;
*Step 2*: Solve the estimation subproblem to obtain $d^{(n)}$ using $P^{(n-1)}$;
*Step 3*: Solve the equilibrium problem to obtain $P^{(n)}$ using $d^{(n)}$ by algorithm SDMSA;
*Step 4*: If convergence criterion is met, then stop; if not, go to step 2.

To solve the VI problem in Step 3 of AUM, a diagonalization method (Sheffi, 1985) is typically applied. When interaction effects are fixed, the VI problem can be turned into an optimization problem. Then the Method of Successive Averages (MSA) can be used to find a local optimum (Powell and Sheffi, 1982), although more efficient methods also exist, e.g. Liu et al. (2009). We propose combining the streamlined diagonalization method and MSA algorithm (SDMSA) for solving *Step 3*.

ALGORITHM SDMSA (Streamlined Diagonalization and MSA)
---
*Step 3.1*: Find an initial hyperpath flow;
*Step 3.2*: At step $n$, load the hyperpath flow onto the network using algorithm HFL in the following; update the transit link costs;
*Step 3.3*: Fix the links costs (not just the interaction effects), then find the descent direction hyperpath flow using algorithm FDD in the following;
*Step 3.4*: Apply a MSA step to update the hyperpath flow;
*Step 3.5*: If convergence criterion is met, then stop, return the proportions ***P*** to upper level; if not, go to step 3.2).

Experiments suggest that the major progress of the lower level assignment model is made at first few iterations. It is natural then to consider returning the lower level proportions $p_{i,h}^w$ *without* convergence to the UE solution in the hope of speeding up AUM. This idea is similar to the idea of *streamlined* diagonalization for solving UE with interactions effects. We call this variant estimator a *double-streamlined* estimator (DSE) since a second "streamline" is used to jump out of the equilibrium subproblem iterations. We call this variant algorithm DSDMSA. This alternative will be used as a benchmark to compare against the proposed algorithm. A significant improvement of SDMSA over DSDMSA would suggest the importance of the equilibrium subproblem.

*3.3 Schedule-Based UE Assignment Model*

The schedule-based UE assignment model in Hamdouch and Lawphongpanich (2008) is adopted to solve Step 3.2.

There are two types of priority rules for transit flows to consider: continuance priority and FCFS priority. Continuance priority means that users already on the vehicle have the priority to continue their service; FCFS priority means that users who wait longer have the priority to board. These are taken into consideration.

The node in the time-expanded network is indexed by $i_h, i \in N, h \in \mathcal{H}$. The TE-network is acyclic and nodes can be arranged in topological & chronological (T&C) order. The tuple $(i_h, j_{h+t_{ij}})$ refers to a link in the TE-network; it is a transit run segment if $i$ and $j$ are different and a waiting link otherwise. Let $z^{s,l,\tau,i_h}$ denote the fractional hyperpath $s \in S_w$ flow appearing at stop $i$ at time $h$ with the arriving route being $l$ and the arriving time being $\tau$. Similarly, let $y^{w,s,l,\tau,a}$ denote the fractional hyperpath $s \in S_w$ flow passing link $a$ with the arriving route being $l$ and the arriving time being $\tau$.

The hyperpath flow loading process in Step 3.2 is summarized by algorithm HFL from Hamdouch et al. (2004). The node flow ***z*** and link flow ***y*** are alternately updated.

The algorithm HFL is listed for completeness.

ALGORITHM HFL (Hyperpath Flow Loading) (Hamdouch et al., 2004)
*Step 3.2.1*: Start with the first node in T&C order;
*Step 3.2.2*: ( $z^{s,l,\tau,i_h}$ update) for node $i_h$, collect flows from all incoming links and departure OD flows; departure flow is described as from "waiting line" and $\tau = h$;
*Step 3.2.3*: ($y^{s,l,\tau,a}$ update for continuance priority flows) load the flows that can and choose to continue to ride; delete saturated links from all preference sets, then set $\tau = 0$;
*Step 3.2.4*: ($y^{s,l,\tau,a}$ update by FIFO rule) for every unsaturated outgoing link, compute the flow representing the first-choice-demand for this link arriving at $\tau$; find the link with the smallest remaining-capacity/first-choice-demand ratio, $\delta$, and this link is saturated first (Marcotte et al., 2004); then each outgoing link has proportion $\delta$ of their first-choice demands fulfilled;
- If $\delta \geq 1$, all flows arriving at $\tau$ are assigned to their first choices; if $\tau = h$ and this is the last node, then stop; if $\tau = h$ and this is not the last node, set current node to be the next node in T&C order, go to step 3.2.2; else $\tau < h$, set $\tau \leftarrow \tau + 1$, then repeat *Step 3.2.4*;
- If $\delta < 1$, update outgoing links' capacity, delete the saturated link from all preference sets; then repeat *Step 3.2.4* with $\tau$ unchanged.

To find the descent direction (Step 3.3 of Algorithm AUM), dynamic programming (DP) is usually applied. Nodes of TE-network arranged in reverse T&C order can be used to label the stages. The decisions to be made are preference sets, $E^{s,l,\tau,i_h}$, which are ordered sets of links ending at stop $i_h$ for users arriving at stop $i$ from bus line $l$ and time $\tau$ on hyperpath $s$. Changes are made from the DP method used in Hamdouch and Lawphongpanich (2008). For example, the preference set is identified by tuple $(s, l, \tau, i_h)$ instead by $(s, \tau, i_h)$ as in Hamdouch and Lawphongpanich (2008). Correspondingly, the state of DP is changed to be $(l, \tau)$ instead of $\tau$. We emphasize that the incoming bus line info $l$ should not be neglected in the identification of the preference set. It's not true that a user can enjoy continuance priority whatever his/her preference set is, as long as his/her transit line can continue. For example, a user arrives at a transfer stop $i$ at time $h$ by a local train $l$ which extends beyond this stop. But if this user decides to alight and wait for an express train, he loses continuance priority. Hence whether a user can enjoy continuance priority is part of the preference set decision; and to make this decision, the incoming line info need to be recorded. In this study, the arrival time information $\tau$ is made to be required for the decision of the optimal value function, instead of being optional as stated in Hamdouch and Lawphongpanich (2008).

The optimal value function $\phi(l, \tau, i_h)$ is the expected cost from $i_h$ to destination. Two classes of users should be distinguished in the calculation of $\phi(l, \tau, i_h)$. The first class of users includes those who arrive at $i_h$ by the waiting line, or by transit lines that have reached their ends. These users have no option of traveling a line with continuance priority. Let $l_a$ be the bus line of an outgoing link $a = (i_h, j_{h+t_{ij}})$. The expected cost $\pi(l, \tau, a)$ of traveling by this outgoing link to a destination is calculated by Eq. (2). If $l_a$ is a waiting line, then the arrival time will not change and stay at $\tau$; otherwise it is changed to $h + t_a$. The user preference set for node $i_h$ is just composed of outgoing links sorted by $\pi(l, \tau, a)$ values. Then $\phi(l, \tau, i_h)$ is the sum of $\pi$'s weighted by the link access probabilities, as shown by Eq. (3). These link-access-probabilities $p(a)$ are found by sending an infinitesimal flow, called virtual flow (Marcotte et al., 2004), from $i_h$ to mix with flows from the loading step. The method to calculate $p(a)$ can be found in Marcotte et al. (2004). As shown above, given the state $(l,\tau)$, the decision at $i_h$ is irrelevant of the upstream decisions; dynamic programming applies here.

$$\pi(l,\tau,a) = \begin{cases} c_a + \phi(l_a, h+t_a, j_{h+t_a}), & \text{for } l_a \text{ being bus line} \\ c_a + \phi(l_a, \tau, j_{h+t_a}), & \text{for } l_a \text{ being waiting line} \end{cases}, \quad (2)$$
$$\text{for each } a = (i_h, j_{h+t_{ij}}) \in \mathcal{A}$$

$$\phi(l,\tau,i_h) = \sum_{a \in \mathcal{A}^+(i_h)} p(a)\pi(l,\tau,a) \quad (3)$$

where
$t_a$: the travel time of this link;
$c_a$: is the travel cost;
$p(a)$: the probability of boarding link $a$;
$\pi(l,\tau,a)$: expected cost of traveling by link $a$ to destination;
$\phi(l,\tau,i_h)$ is the expected cost from $i_h$ to destination;
$\mathcal{A}$: set of arcs;
$\mathcal{A}^+(i_h)$: set of arcs emanating from node $i_h$.

The second class of users have the option to continue. Hence they need to make a decision of whether to continue to ride or alight. For the former case, the preference set just consist of $l$, and the expected cost $\phi_{continue}$ is calculated by Eq. (4). For the latter case, the preference set is composed of outgoing links sorted by $\pi$ values as before and the expected cost $\phi_{alight}$ is also calculated by Eq. (2)-(3). Note what is implied by Eq. (3) in this case is that once a user alights his or her line, he or she needs to queue to board any line even if that line is the line where he or she was from. Finally, the expected cost $\phi(l,\tau,i_h)$ is given by Eq. (5). And the preference set is decided correspondingly. The steps for finding the descent direction, with modifications described above, are listed in algorithm FDD for completeness.

$$\phi_{continue}(l,\tau,i_h) = \pi^{l,\tau}(a) = c_a + \phi(l, h+t_a, j_{h+t_a}), \quad l_a = l \quad (4)$$

$$\phi(l,\tau,i_h) = \min\{\phi_{continue}(l,\tau,i_h), \phi_{alight}(l,\tau,i_h)\} \quad (5)$$

ALGORITHM FDD (Finding Descent Direction)

*Step 3.3.1*: Start with the first destination, $r$;
*Step 3.3.2*: Start with the first node in reverse T&C order; if it's origin node ($r_h$), then the distance $\pi$ is zero; otherwise, set $\pi$ to infinity;
*Step 3.3.3*: At current node $i_h$, compute the cost $\phi$ of boarding each outgoing link for each arriving time $\tau$;
*Step 3.3.4*: For all possible states $(l,\tau)$, choose the corresponding way of finding the preference sets and optimal value function $\pi$;
*Step 3.3.5*: If $i_h$ is the last node in reverse T&C order and $r$ is the last destination, then stop; if $i_h$ is the last node and $r$ is not the last destination, then $r \leftarrow r+1$, go to step 3.3.2; otherwise, set current node to be the next node, go to step 3.3.3.

The initial strategy for each destination $r$ can be found based on the nodes' distances to $r$ under "free flow" condition, where "free flow" means no congestion effect or capacity constraint. At node $i_h$, all downstream nodes are sorted according to their distances to obtain the preference set. It's also recommended to include all these links with finite distances into the preference sets to help ensure feasibility.

In summary, the schedule-based passenger flow estimation method is illustrated in a data flow diagram in Figure 2.

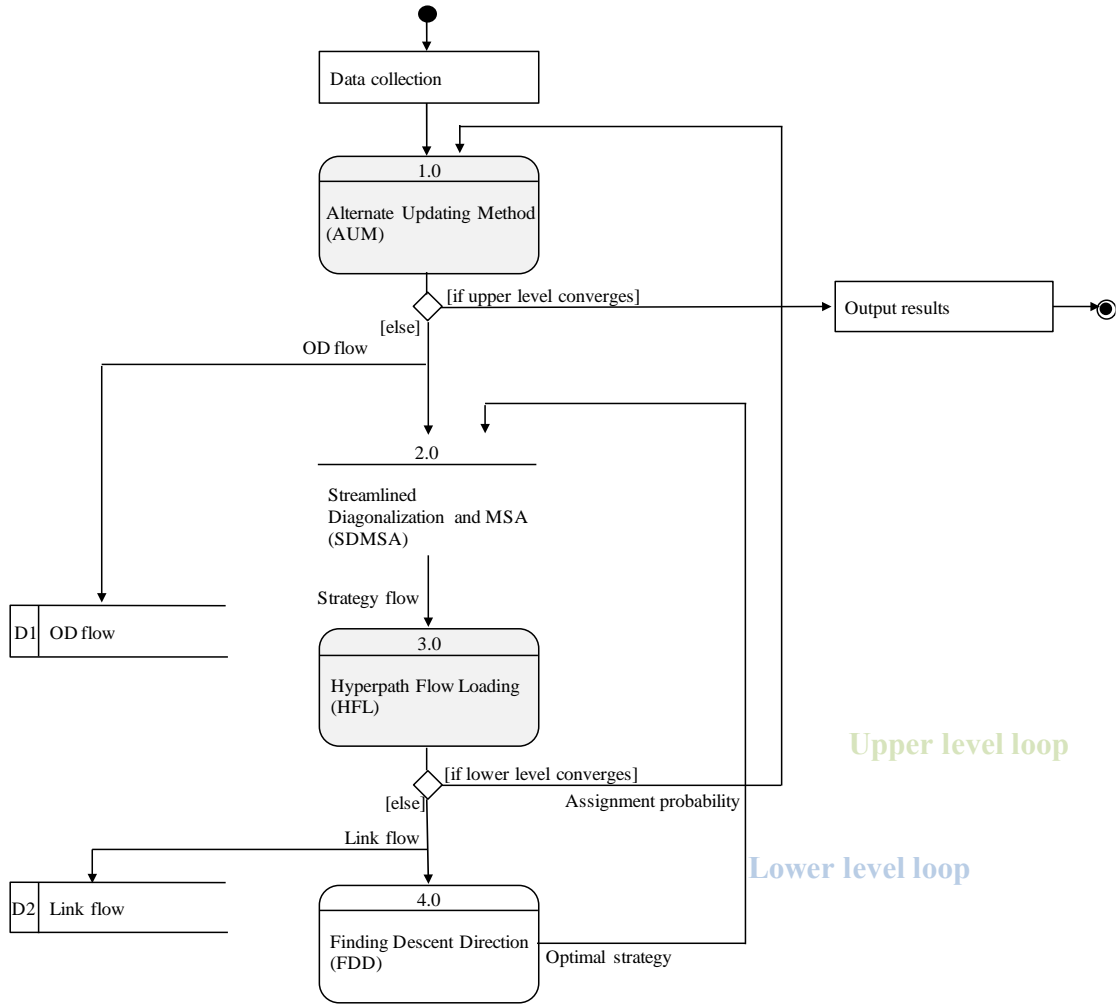

Notes: SDMSA, HFL, FDD are in fact sub-steps of the Alternate Updating Method (AUM).

Figure 2. Data flow diagram of the proposed estimation method considering new (white) and existing (gray) algorithms from literature.

## 3.4 Solution Existence, Uniqueness and Algorithm Convergence

The model's goal is to minimize a quadratic function, bounded below by zero; the feasible set of the lower level model is non-empty, closed and bounded; hence the optimal solution of the MPEC always exists. Generally, an equilibrium in this problem is not guaranteed to be unique so the model is nonconvex. As such, our algorithm obtains a local optimum.

Relative gap is usually used for the convergence test of the lower level VI problem (Marcotte et al., 2004), as shown in Eq. (6). $C(f)$ refers to the total cost of the current loaded strategy flows $f$; it can be obtained immediately after each loading step, where we have a vector $f = [f_1, f_2, \ldots, f_n, 0 \ldots, 0]$ after $n$ MSA steps. For the estimation subproblem, the average relative change (ARE) (Sheffi, 1985) or maximum relative change (Yang, 1995) can be used for variables of interest like OD flows or link flows. When the OD flow variables are chosen, ARE is calculated using Eq. (7). The problem with ARE is that it can only be calculated when the denominator is non-zero. Mean squared error (MSE) can be used instead, as shown in Eq. (8).

$$\frac{|C(f) - C(g)|}{C(f)} \leq \epsilon \tag{6}$$

$$\frac{1}{W} \sum_{w \in \mathcal{W}} \frac{|d^{w(n+1)} - d^{w(n)}|}{d^{w(n)}} \leq \epsilon' \tag{7}$$

$$\frac{1}{W} \sum_{w \in \mathcal{W}} (d^{w(n+1)} - d^{w(n)})^2 \leq \epsilon'' \tag{8}$$

### *3.5 Computational and Statistical Considerations*

The first consideration is the unit time interval. It should be small enough that all transit segment travel times and waiting times be expressed as integral multiples of this interval such that the TE-network is applicable. Passengers and vehicles arriving within the same interval are considered happening simultaneously. Passengers can transfer between simultaneously arriving vehicles; simultaneously arriving users have the same FIFO priority. To satisfy these requirements, the unit time interval is expected to be 1 minute or even less.

The second consideration is about horizon $H$ setting. It is good practice to reserve 30 to 60 minutes ahead of the targeted period for initial loading. Another 30 to 60 minutes trailing time is also desired, since passengers during this time may not be able to get to their destinations. As a result, the horizon would be 2 hours or longer; it means 120 units if 1 minute is selected as the unit interval.

**Proposition 1.** *The space complexity and time complexity for the estimation algorithm AUM are both $O(N^2 L)$.*

**Proof**. First we discuss the space complexity. At first glance, the critical part is storing the node flow variable $z^{s,l,\tau,i_h}$ and link flow variable $y^{s,k,l,\tau,a}$; these two variables are in fact 7-dimensional $(q, r, h, s_r, l, \tau, i_h)$. The number of nodes in TE-network is $NT$. $z$ has dimension of $(NH)^3 S(L+1)$, where $S$ is the maximum number of hyperpaths for each $w$. The TE-network is typically sparse, just like the physical network, so the number of links, $A$, is comparable to the number of nodes, namely $A \approx NH$. Then $y$ has the same order of magnitude as $z$. However, there is no need to store the whole $z$ and $y$ vector; storing $y$ and $z$ would no longer be critical. Instead, the preference set takes up the most storage. It's 5-dimensional - $(r, s_r, l, \tau, i_h)$ and its dimension is $N^2 S(L+H)H$, where $S$ and $H$ are model parameters. Hence the space complexity is $O(N^2 L)$. The computational time is proportional to the total number of stages $N^2 H$ and the time spent at each stage; the latter is bounded by $SL$. the average time spent at each stage is about $SL/2$. Hence computation time would also increase by order $O(N^2 L)$. ∎

Experimental results show that a computer with 128GB RAM and 2 quad-core CPU takes about 40 minutes to solve each lower level assignment iteration in Python for a network with 60 stops, 5 transit routes and 120 min horizon, and tolerance being 0.005 for both subproblems. In most previous studies with schedule-based assignment, the models were applied to very small networks, say 5 nodes in Marcotte et al. (2004) and 6 nodes network with 6 time intervals in Hamdouch and Lawphongpanich (2008). Nuzzolo et al. (2012) applied their schedule-based assignment model to a network with 11 traffic zones and 9 transit lines; their example has a horizon of 2 hour divided into 5-min slices. Their assignment takes one hour on a personal computer (Intel Core 2 Quad CPU,4 Gb RAM), but it was not an estimation model. Computation time for our algorithm can be improved significantly for implementation using more efficient languages like C++ and relaxing the tolerance to 1%, which Hamdouch and Lawphongpanich (2008) noted to be sufficient. In spite of this, the computational time for large network (with ~500 stops) would still be formidable. We argue that schedule-based behaviors mostly appear in transit systems with infrequent services, which typically lies in small cities with modest size.

The number of stops measured affects the error of the estimator. The source of error includes: 1) users may not strictly follow the assumptions of assignment; 2) measurements are liable to errors. These factors are subject to randomness. The estimator variance of an over-determined problem usually has this relation:

$$\text{estimator variance} = \text{population distribution variance} / \text{sample size}$$

However, the flow estimation problem is under-determined; the estimation error is difficult to formulate. We get an upper bound for the MSE instead.

**Proposition 2**. *The upper bound for the MSE of this simpler problem is bounded by a linear function of the number of measurements.*

**Proof**. Let $m$ be the number of measurements. The flows are bounded below by zero and above by some reasonable upper bound, say capacity $u$. Hence the errors are also bounded, even if there are no measurements at all. Let's study a simpler linear problem instead:

$$Ax = b$$
$$0 \leq x \leq u$$

where $x$ is a $n$ dimensional vector to estimate; $A$ is a $n$ by $m$ coefficient matrix and $n \geq m$ (under-determined); $b$ is a random vector affected by error $\varepsilon$.

If there are no bounds on $x$, we typically solve this by $x = A^T(AA^T)^{-1}\boldsymbol{b}$, where $A^T(AA^T)^{-1}$ is the pseudo-inverse; this is the solution with least $L^2$-norm. Here we still solve this problem by finding the least $L^2$-norm solution. After changing of variables $x = P\tilde{x}$, we can transform the coefficient matrix $A$ to column echelon form $AP = [M \vdots 0]$; then the original problem becomes:

$$[M \vdots 0]\tilde{x} = b$$
$$0 \leq \tilde{x} \leq P^{-1}u$$

Suppose $M$ is a non-degenerate $m$ by $m$ matrix. Let $\bar{u}$ be the largest component of $P^{-1}u$. The largest error of the least $L^2$-norm solution happens when the real value of the $\tilde{x}_i$'s corresponding to the zero column takes on the upper bound value (elements of $P^{-1}u$). The MSE of $\tilde{x}$ satisfies Eq. (9).

$$\begin{aligned}
MSE &\leq \frac{1}{n}[(n-m)\bar{u}^2 + \mathbf{1}^T var(M^{-1}\boldsymbol{b})] \\
&= [\bar{u}^2 + \frac{\mathbf{1}^T var(M^{-1}\boldsymbol{b})}{n}] - \frac{\bar{u}^2}{n}m \\
&= [\bar{u}^2 + \frac{\mathbf{1}^T M^{-1} var(\varepsilon) M}{n}] - \frac{\bar{u}^2}{n}m \\
&= (u^2 + \sigma^2) - \frac{\bar{u}^2}{n}m \\
&= c_1 - c_2 m
\end{aligned} \quad (9)$$

where we assume that the errors are *i.i.d.* with variance $\sigma^2$. The total squared error of $x$ is the same with $\tilde{x}$ if we assume that the change of variables is an orthogonal transformation; the MSE of $x$ also follows Eq. (9). Hence the MSE of this simpler problem is bounded by a linear function of the number of measurements $m$. We expect our much more complex non-linear under-determined flow estimation problem has a similar form of error bound. ∎

## 4 COMPUTATIONAL EXPERIMENTS

Several computational experiments are conducted to verify the assignment algorithm, the estimation algorithm, and to validate the methodology using real data. The source code for models and data for examples are shared at BUILT@NYU (https://github.com/BUILTNYU/transit-flow-estimation/tree/master/schedule-based).

### *4.1 Dynamic UE Assignment Illustration*
The example in this section is used to verify the schedule-based UE assignment algorithm. The transit network is shown in Figure 3. There are two bus lines *l1* and *l2*. Vehicle capacity is 200 for both lines. There are 100 users for OD *(N1,N2,0)* and 150 users for *(N5,N2,0)*. Suppose that waiting cost equals (1+ δ) times of travel time, where δ is a small positive number; and assume that congestion effect exists. Hence users *N1 → N2* would rather wait at *N3* for *l2* than detour on a congested bus to *N4* or *N5*. The 150 *N5 → N2* users would always take *l2* line.

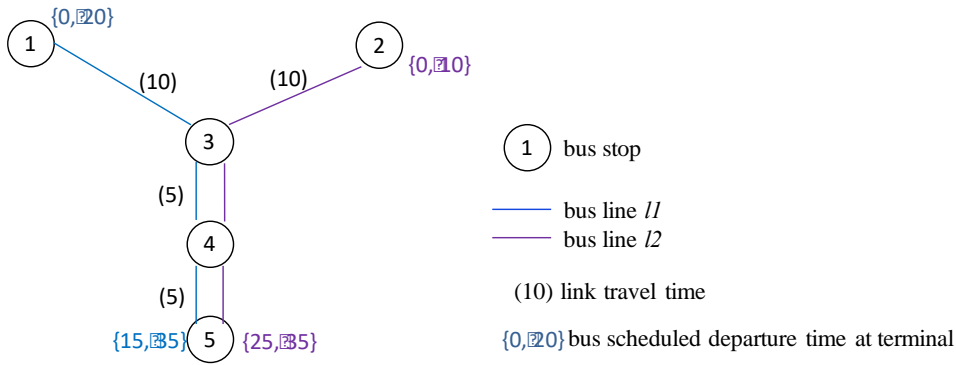

Figure 3. Transit network in example 1.

In the first iteration, 100 units of flow from *N1* take *l1* to *N3*, then wait until time 35 when the first *l2* run pass by *N3*. But there are only 50 units of remaining capacity, the probability of boarding is 0.5. The rest wait for the next *l2* run that arrives 10 min later. The optimal strategy is to divert *N1* originated users to transfer at node *N4* for continuance priority at *N3*. After one MSA step, 50 *N1* originated users follow the initial strategy, and the other 50 users transfer at node *N4*.

To find the optimal strategy for the second iteration, we can see that at node *N4* the virtual flow's probability of boarding the first *l2* run is still 1.0, since 50 users from *N1* plus 150 users from *N2* equals the capacity 200 and the virtual flow is infinitesimal. Hence the optimal strategy does not change.

In the third iteration, 2/3 of 100 *N1* originated users are diverted to *N4* and the probability of boarding the first *l2* run becomes smaller than 1. The optimal strategy now diverts *N1* users to transfer at further stop upstream, *N5*. The steps are shown in Figure 4. Users would detour upstream further and further to gain continuance priority to bring down their own costs, until there is no benefit to the action. This phenomenon is called a "detouring upstream phenomenon" and is expected to happen in general schedule-based UE assignment models.

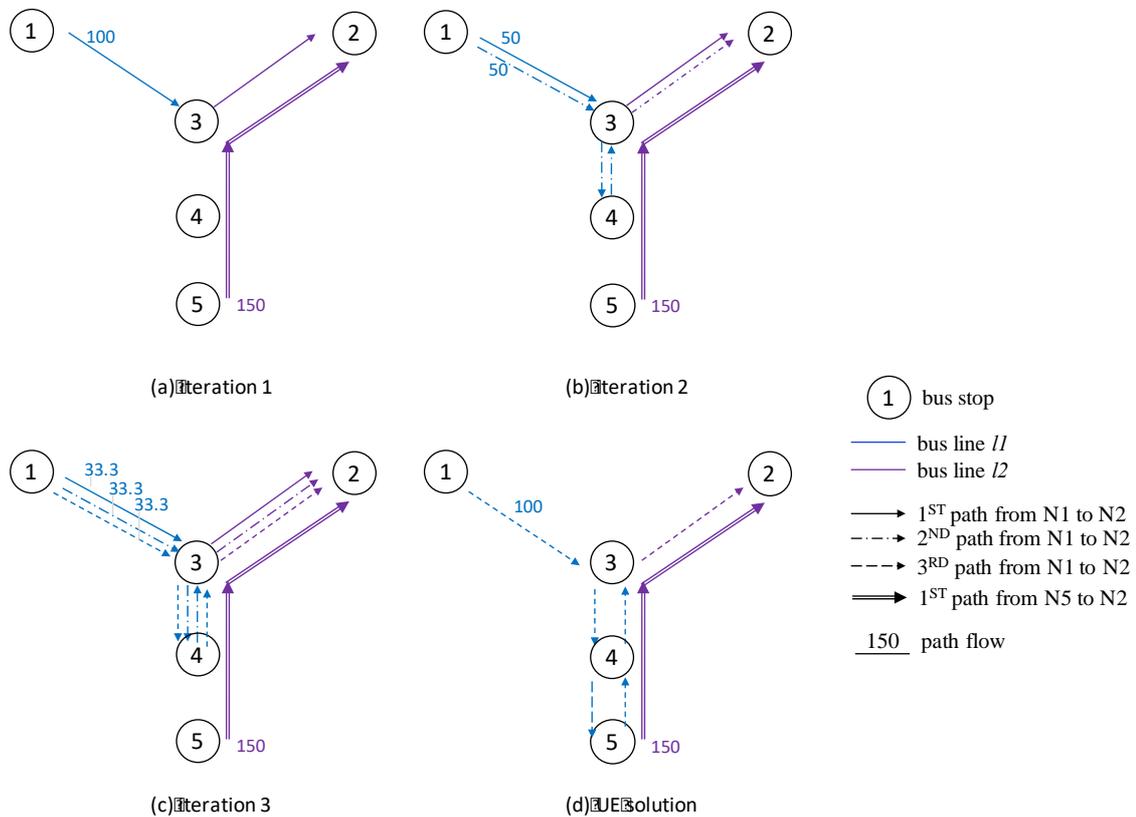

Figure 4. Path flows realized from hyperpath flows over iterations; (a) 150 units N5→N2 flow choose path 5-4-3-2; 100 units N1→N2 flow choose path 1-3-2, half of which cannot board the first bus run at stop 3; (b) half of N1→N2 flow divert to path 1-3-4-3-2 to seize continuance priority over the rest half; (c) 2/3 of N1→N2 flow divert to path 1-3-4-3-2; (d) all N1→N2 flow divert to path 1-3-4-5-4-3-2 at equilibrium.

At the $n$-th iteration ($n > 3$), $1/n$ proportion of *N1* originated users transfer at *N3*; $2/n$ proportion of *N1* originated users transfer at *N4*; the rest of the $(n-3)/n$ proportion of the *N1* users transfer at node *N5*. The UE solution in this simple example is letting almost all *N1* users transfer at node *N5*. The total cost of all flows is $C(f) = 50(55) + (50 + 150)45 = 11750$. At iteration $n > 3$, the total cost of the users' optimal strategy is $C(g) = 100(p)(45) + 100(1-p)(55) + 150(45)$, where $p = 50/(100(n+3)/3)$ is the probability of boarding the first run of *l2* at *N3*. The relative gap is shown to be $|C(g) - C(f)|/C(f) = 30/(235(n-3))$. This gap decays at a speed of $O(n^{-1})$, i.e. the convergence rate is constant, although this is not guaranteed for MSA.

### *4.2 OD Demand and Ridership Estimation Verification*
This section tests the effectiveness of the estimation model to retrieve the OD flows and ridership given that the transit flows are indeed distributed according to a schedule-based UE assignment with some controlled noise.

The network used in this example is the Sioux Falls network shown in Figure 5 with three transit lines indicated in blue, green, and red. There are 14 stops on the bus lines. The time interval is 1 min and the horizon is 120 min. The congestion function is assumed to be of quadratic form: $c = [(V/C)^2 + 1]t$, $0 \leq V \leq C$, where $V$ is the run flow; $C$ is the capacity and $t$ is the travel time. This function is convex; it takes on 1.0 at v/c = 0 and 2.0 at v/c=1.0. The vehicle capacity is 100. The bus headways are 15 minutes. The detailed settings can be found in the 4-2 example in the Github link provided earlier.

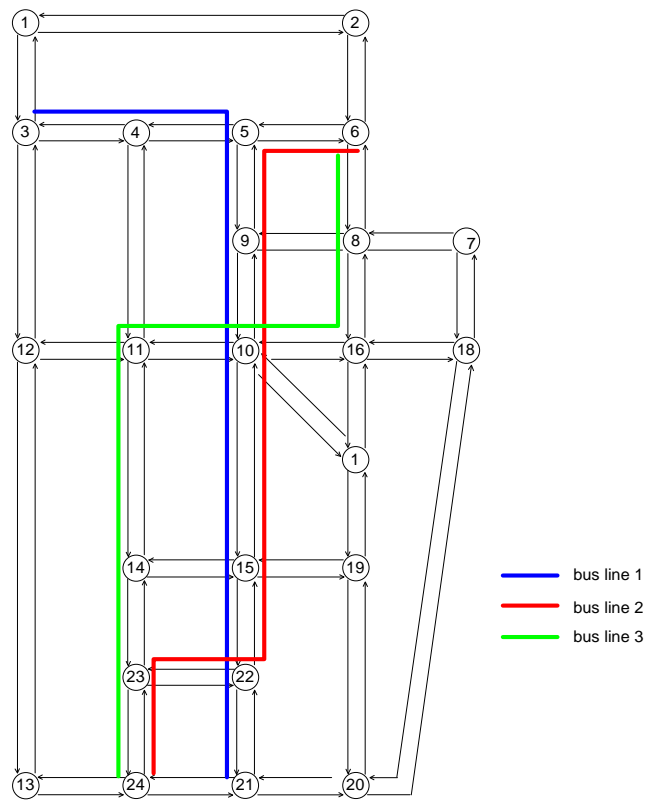

\* All bus lines are bi-directional, with 15-minutes headway and 100-units capacity.

Figure 5. Bus network added to Sioux Falls.

The measurement period is assumed to be the same as the time unit. The results represent the best possible estimation based only on the stop-level aggregated counting data. This establishes the upper bound for the performance of the proposed model. The demands are also loaded onto the network using the same schedule-based assignment model assumed by the estimation model. The "real" link volumes and "observed" measurement counts can be found in the 4-2 example in Github. The equilibrium subproblem's threshold $\varepsilon$ is set to be 0.005. With this tolerance and software and hardware, each equilibrium subproblem iteration takes less than 5 minutes; a single estimation subproblem iteration takes about 30 minutes. We ran it up to 500 iterations without any stopping condition to observe the behavior of the algorithm.

The demands are chosen to be spatially and temporally heterogeneous; the performances of estimators are shown in Figure 6. The trajectory of the objective value in Eq. (1a) (SSE) is shown in Figure 7. In both examples the convergence is not monotone. Furthermore, the performance of DSDMSA is clearly not satisfactory. The MSE of the DSDMSA is significantly larger than that of the standard one. SDMSA converges after 350 iterations; meanwhile, DSDMSA shows no sign of convergence. Therefore, the double-streamlined estimator is not recommended for practical use.

The details of the estimation results using SDMSA are shown from Table 2 to Table 4. The estimation for minute OD demand ($N5 \rightarrow N15$) are displayed in Table 2. The estimation results for hourly OD demands (*N4, N5* originated) are shown in Table 3. Table 4 shows the transit run ridership estimation for south bound line *l1*. "Real" refers to synthesized real data. MSE of SDMSA is 0.076 for the minute-level OD demand, 156.7 for hourly OD demand and 11.6 for ridership. The results are satisfactory. As $N$ increases the number of model variables grows at a rate $O(N^2 H)$ while the number of measurements grows at rate $O(NH)$; this estimation problem would become more and more underspecified.

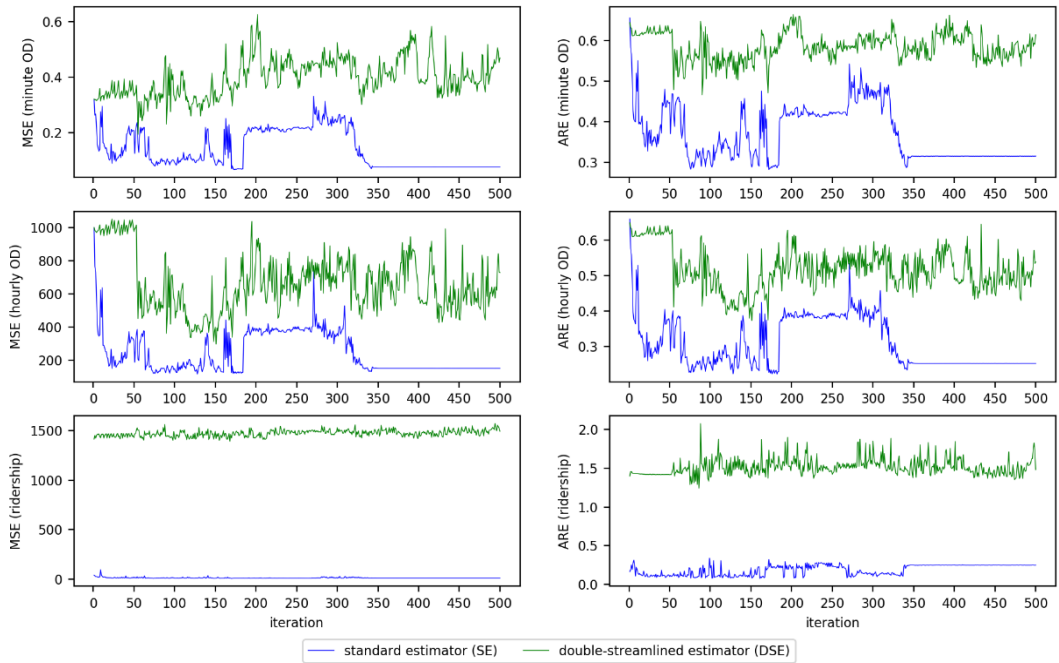

Figure 6. Performance of estimators over iterations.

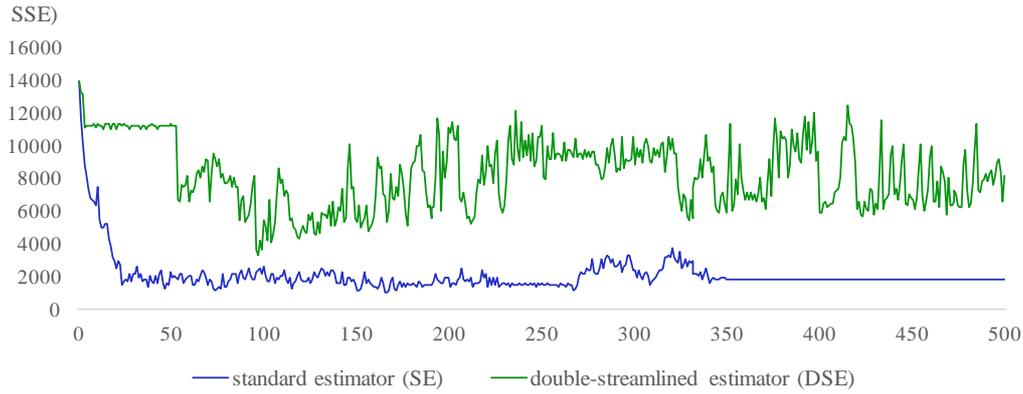

Figure 7. Upper-level goal value over iterations.

Table 2. Minute OD demands estimation for pair *N5-N15*.

| h | Real | SE | h | Real | SE | h | Real | SE | h | Real | SE | h | Real | SE |
|---|---|---|---|---|---|---|---|---|---|---|---|---|---|---|
| 0 | 4 | 3 | 12 | 4 | 2 | 24 | 8 | 9 | 36 | 8 | 7 | 48 | 8 | 6 |
| 1 | 4 | 3 | 13 | 4 | 2 | 25 | 8 | 9 | 37 | 8 | 7 | 49 | 8 | 6 |
| 2 | 4 | 3 | 14 | 4 | 2 | 26 | 8 | 9 | 38 | 8 | 7 | 50 | 8 | 6 |
| 3 | 4 | 3 | 15 | 4 | 2 | 27 | 8 | 9 | 39 | 8 | 7 | 51 | 8 | 6 |
| 4 | 4 | 3 | 16 | 4 | 2 | 28 | 8 | 9 | 40 | 8 | 7 | 52 | 8 | 6 |
| 5 | 4 | 3 | 17 | 4 | 2 | 29 | 8 | 9 | 41 | 8 | 7 | 53 | 8 | 6 |
| 6 | 4 | 2 | 18 | 4 | 2 | 30 | 8 | 9 | 42 | 8 | 7 | 54 | 8 | 6 |
| 7 | 4 | 2 | 19 | 4 | 2 | 31 | 8 | 9 | 43 | 8 | 7 | 55 | 8 | 6 |
| 8 | 4 | 2 | 20 | 8 | 4 | 32 | 8 | 9 | 44 | 8 | 7 | 56 | 8 | 6 |
| 9 | 4 | 2 | 21 | 8 | 9 | 33 | 8 | 9 | 45 | 8 | 7 | 57 | 8 | 6 |
| 10 | 4 | 2 | 22 | 8 | 9 | 34 | 8 | 9 | 46 | 8 | 6 | 58 | 8 | 6 |
| 11 | 4 | 2 | 23 | 8 | 9 | 35 | 8 | 8 | 47 | 8 | 6 | 59 | 8 | 6 |

Table 3. Hourly OD demands estimation results for origins *N4, N5*.

| q | r | Real | SE | q | r | Real | SE | q | r | Real | SE |
|---|---|---|---|---|---|---|---|---|---|---|---|
| N5 | N4 | 0 | 2 | N5 | N21 | 400 | 334 | N6 | N11 | 0 | 9 |
| N5 | N5 | 0 | 0 | N5 | N22 | 0 | 0 | N6 | N14 | 0 | 4 |
| N5 | N6 | 0 | 2 | N5 | N23 | 0 | 47 | N6 | N15 | 0 | 6 |
| N5 | N8 | 0 | 3 | N5 | N24 | 0 | 1 | N6 | N16 | 0 | 20 |
| N5 | N9 | 0 | 1 | N6 | N4 | 0 | 0 | N6 | N21 | 0 | 54 |
| N5 | N10 | 0 | 1 | N6 | N5 | 0 | 1 | N6 | N22 | 0 | 20 |
| N5 | N11 | 0 | 1 | N6 | N6 | 0 | 0 | N6 | N23 | 400 | 264 |
| N5 | N14 | 0 | 2 | N6 | N8 | 0 | 1 | N6 | N24 | 0 | 2 |
| N5 | N15 | 0 | 1 | N6 | N9 | 0 | 0 | | | | |
| N5 | N16 | 0 | 2 | N6 | N10 | 0 | 9 | | | | |

The performances of the model under different measurement periods are tested and the results are shown in Figure 8. As the period becomes larger, the estimation errors for minute OD and ridership increase rapidly, although the hourly OD estimation does not seem to be influenced much. The error of ridership estimation soars when the period gets larger than 15 min. This suggests that the measurement period should be 15 min or less.

The estimation errors under different numbers of stops measured are shown in Figure 9. The robustness of this estimation model is tested by introducing a perception error to weaken the assumption of route choice behaviour made in the estimation model. Travelers in the real world may not necessarily travel exactly as predicted under the schedule-based UE model. The artificial perception errors (percentage) are assumed to be log-normally distributed. Multinomial Logit is used to assign users to hyperpaths: $Pr(hyperpath = i) = e^{-\theta c_i} / \sum_j e^{-\theta c_j}$, where $\theta$ is 0.1; $c_i$ is expected hyperpath travel cost (min). The results are shown in Figure 10. The x-axis is the variance of log(percentage noise).

Table 4. Transit run ridership estimation for south bound line *l1*.

| start | end | Real | SE | start | end | Real | SE | start | end | Real | SE |
|---|---|---|---|---|---|---|---|---|---|---|---|
| N4_0 | N5_5 | 0 | 0 | N10_45 | N15_50 | 100 | 100 | N4_75 | N5_80 | 0 | 0 |
| N5_5 | N9_10 | 11 | 11 | N15_50 | N22_55 | 100 | 100 | N5_80 | N9_85 | 0 | 3 |
| N9_10 | N10_15 | 11 | 11 | N22_55 | N21_60 | 100 | 100 | N9_85 | N10_90 | 0 | 3 |
| N10_15 | N15_20 | 11 | 11 | N4_45 | N5_50 | 0 | 0 | N10_90 | N15_95 | 0 | 3 |
| N15_20 | N22_25 | 11 | 11 | N5_50 | N9_55 | 80 | 77 | N15_95 | N22_100 | 7 | 3 |
| N22_25 | N21_30 | 24 | 24 | N9_55 | N10_60 | 80 | 76 | N22_100 | N21_105 | 12 | 30 |
| N4_15 | N5_20 | 0 | 0 | N10_60 | N15_65 | 100 | 85 | N4_90 | N5_95 | 0 | 0 |
| N5_20 | N9_25 | 46 | 46 | N15_65 | N22_70 | 100 | 100 | N5_95 | N9_100 | 0 | 0 |
| N9_25 | N10_30 | 46 | 46 | N22_70 | N21_75 | 100 | 100 | N9_100 | N10_105 | 0 | 0 |
| N10_30 | N15_35 | 46 | 46 | N4_60 | N5_65 | 0 | 0 | N10_105 | N15_110 | 0 | 0 |
| N15_35 | N22_40 | 46 | 46 | N5_65 | N9_70 | 61 | 60 | N15_110 | N22_115 | 0 | 0 |
| N22_40 | N21_45 | 64 | 64 | N9_70 | N10_75 | 61 | 60 | N4_105 | N5_110 | 0 | 0 |
| N4_30 | N5_35 | 0 | 0 | N10_75 | N15_80 | 61 | 60 | N5_110 | N9_115 | 0 | 0 |
| N5_35 | N9_40 | 84 | 100 | N15_80 | N22_85 | 73 | 70 | | | | |
| N9_40 | N10_45 | 100 | 100 | N22_85 | N21_90 | 100 | 100 | | | | |

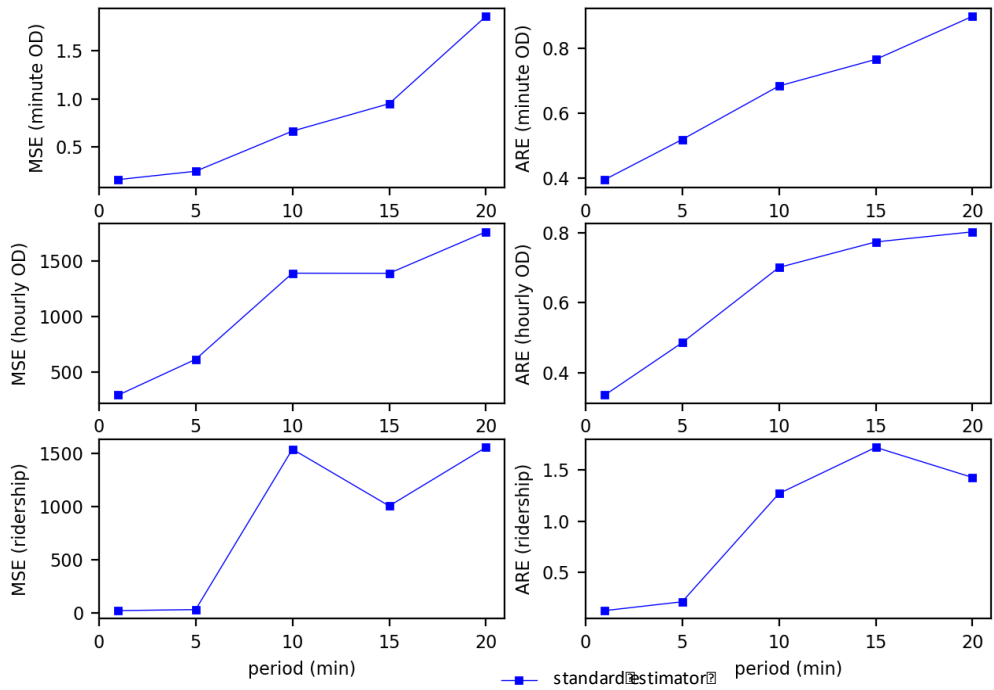

Figure 8. Performance of standard estimator under varying measurement periods.

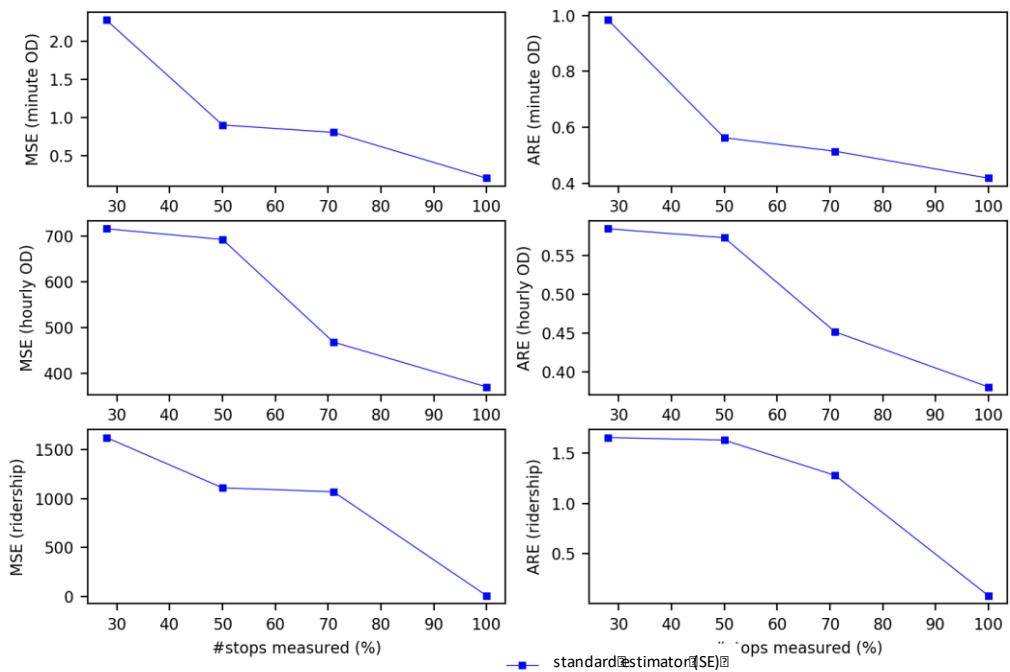

Figure 9. Performance of standard estimator under varying number of stops measured.

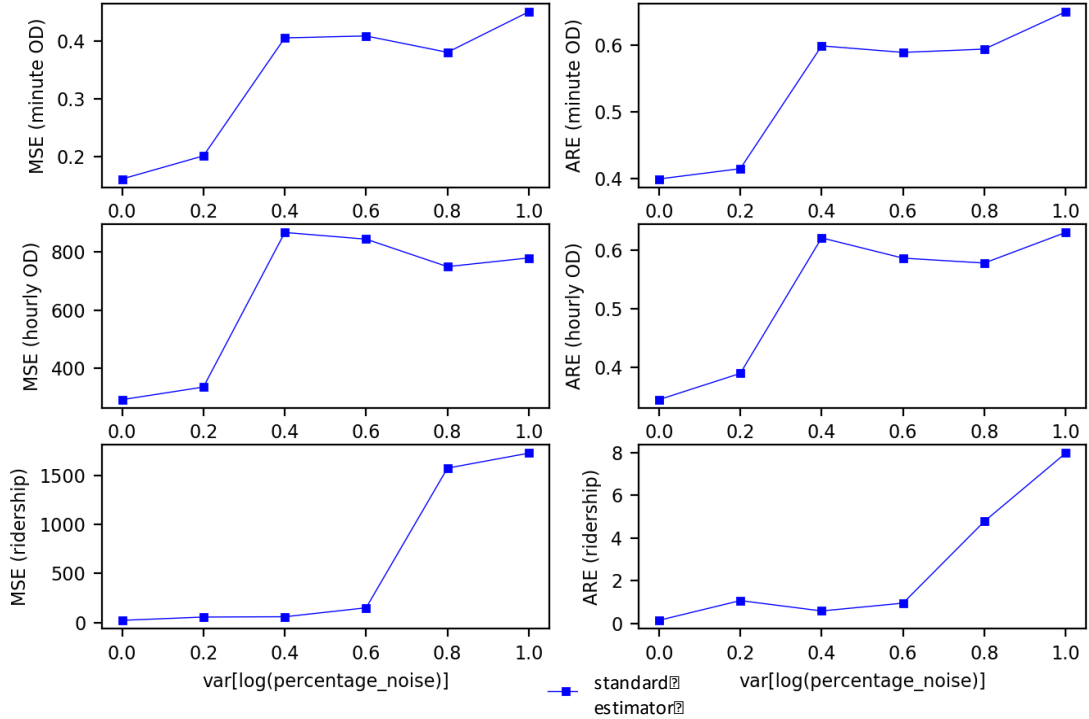

Figure 10. Standard Estimator Performances under different noise levels.

When the variance is within 0.25, hourly OD demands and ridership estimation MSE lies within 20%. This result suggests that the estimation model is effective when the perception errors are within the interval $(e^{-\sqrt{1/4}}, e^{\sqrt{1/4}}) \approx (60\%, 160\%)$ with confidence 0.68 for this particular example.

## 5 MODEL VALIDATION WITH SHANGHAI BUS DATA

### 5.1 Data for base scenario

We implement the model on a local bus network in Qingpu District, Shanghai, China to validate the estimation model and to show how it can be applied to transit operations. There are 4 bus lines, 120 segments, 55 bus stops, on this subnetwork; the bus routes are shown in Figure 11. The headways are between 5 to 10 minutes in the peak hour and between 10 to 20 minutes off peak. When we consider one-minute time intervals across 60 minutes, this makes the problem significantly larger in size than any prior study of schedule-based transit assignment (which were limited to 6 nodes with 6 time intervals), and the first schedule-based passenger flow estimator based on a real data set.

We have 10GB of individual level swipe data upon boarding collected from the network obtained between July 2016 to September 2016 from a Shanghai open data challenge (SODA, 2016). For that year, there were 15 million records for modes including bus, metro, and ferry. Out of that set, 5 million records are bus trips and 50,000 of those correspond to the Qingpu bus network evaluated in this case study. The recorded fields are card Id, date, boarding time, line, mode, cost and discount type. The stop info is not recorded for bus, but it can be partly extracted using the second-by-second boarding time data. User OD demands are then inferred by combining AM and PM boarding records, after which they are loaded onto transit segment flows by aggregating transit run flows based on their timestamps. For this experiment we assume those inferred ODs are the ground truth corresponding to the AM peak hour (7:00-8:00AM) of Friday, July 1, 2016. Total AM peak hour demand is about 1,400. The inferred OD demands are illustrated by Figure 12.

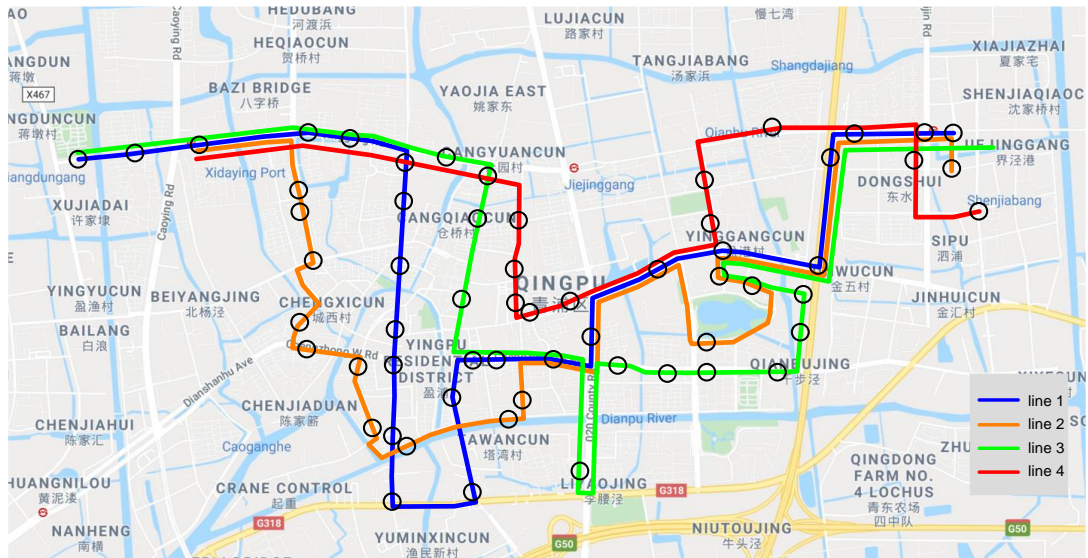

Figure 11. Bus lines at Qingpu District, Shanghai, China.

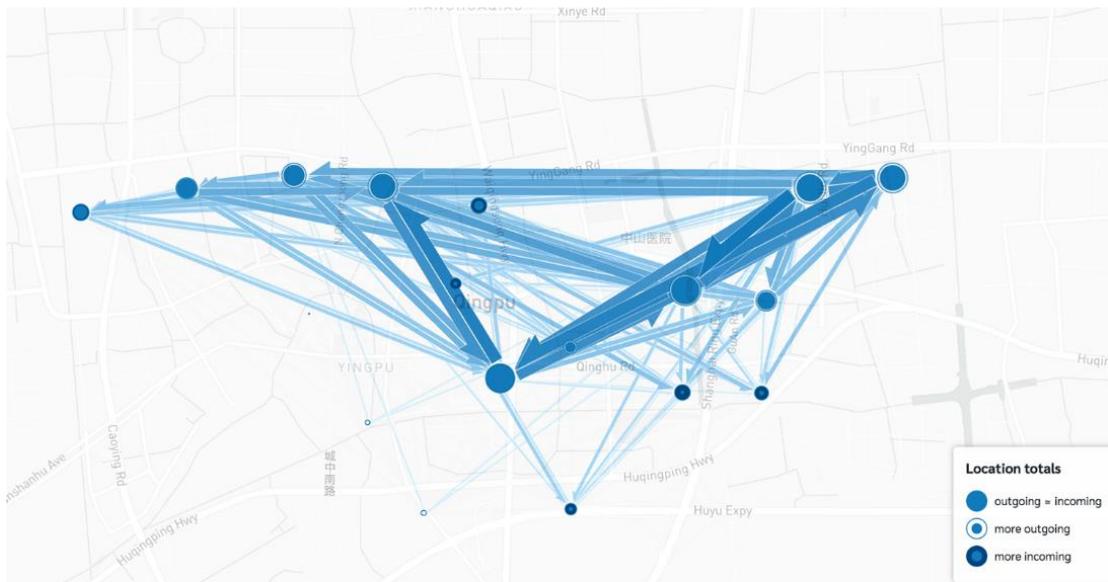

Figure 12. 7:00-8:00AM OD flows in Qingpu.

*5.2 Validation of estimation model for base scenario*

We use standard error and average relative error as the performance indices as shown in Eq. (8) – (10). Both the upper level and lower level thresholds $\varepsilon$ and $\varepsilon''$ are set to 0.005. Using those tolerances, each lower level iteration takes about 38 minutes to run; a single upper level takes about 6 hours on average; the estimation algorithm converged after 75 hours. Since the code was written in the notoriously slow Python, it is possible to improve its run time for online implementation. For example, the algorithm can be made faster if JIT compiling is applied using PyPy or Numba or rewritten in Cython or C++. Increasing the tolerance to 0.01, as noted to be sufficient by Hamdouch and Lawphongpanich (2008), should cut the run time alone by a significant chunk. We can further reduce computational time by using 5-minute intervals instead of 1-minute intervals, which should then make the model operable online.

$$\bar{\mu} = \frac{1}{\#Seg} \sum_{(i,j)\in \mathcal{N}\times\mathcal{N}} \sum_{h\in\mathcal{T}} y^{s,l,\tau,(i_h,j_{h+t_{ij}})} \tag{10}$$

$$std\ error = \left(\frac{1}{\#Seg} \sum_{(i,j)\in \mathcal{N}\times\mathcal{N}} \left(\sum_{h\in\mathcal{T}} y^{s,l,\tau,(i_h,j_{h+t_{ij}})} - f_{(i,j)}\right)^2\right)^{1/2} \tag{11}$$

$$ARE = \frac{1}{\#Seg}\left(\left|\sum_{h\in\mathcal{T}} y^{s,l,\tau,(i_h,j_{h+t_{ij}})} - f_{(i,j)}\right|/f_{(i,j)}\right) \tag{12}$$

where $f_{(i,j)}$ is the inferred segment flow for segment $(i,j)$; $\#Seg$ is the number of transit segments on the transit network.

The convergence trajectory is shown in Figure 13. The algorithm is clearly convergent.

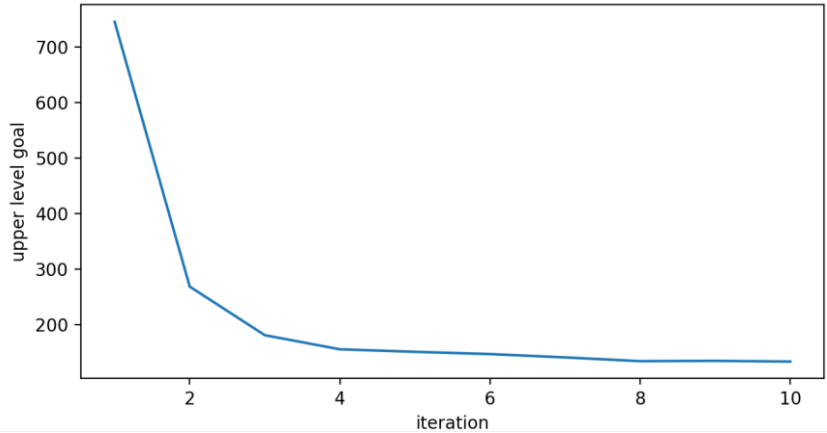

Figure 13. Upper-level objective value over iterations.

The test results are shown in Table 5 and visualized in Figure 14. The "inferred flow" in (a) refers the flows observed and inferred from the data; "estimated flow" in (b) refers to the flow estimated using the proposed algorithm with the aggregated count data. The difference between the observed means and the estimated means are quite close, with the average of the segment flows only 2.6% off from the average of the observed flows. The standard error (0.06) and relative error (72.9%) are somewhat larger for minute-level OD flow but they decrease substantially when the flows are aggregated to hourly OD flows and distributed to segment flows; the relative error of the segment flow is 42.5%. By comparison, the case study of Irvine network by Bierlaire and Crittin (2004) had relative mean errors in the range of 11% to 30% while Cascetta et al. (1993) had a case study of dynamic OD estimation for the Brescia-Verona-Vicenza-Padua motorway in Italy with 19 centroids, 54 links, and 171 OD pairs, having relative mean errors of 17% to 49%. Considering our study considers a more complex transit schedule-based assignment constraint, this is satisfactory at the segment flow level. The visualization of the observed and estimated flows shown in Figure 14 also suggest that the major segments are well-estimated.

Table 5. Estimator performance compared to inferred flow

|  | Observed mean $\mu$ | Estimated mean $\bar{\mu}$ | Std error (person) | ARE (%) |
|---|---|---|---|---|
| Minute OD flow | 0.05 | 0.06 | 0.06 | 72.9% |
| Hourly OD flow | 3.26 | 3.39 | 1.98 | 58.4% |
| Segment flow | 92.5 | 94.9 | 33.1 | 42.50% |

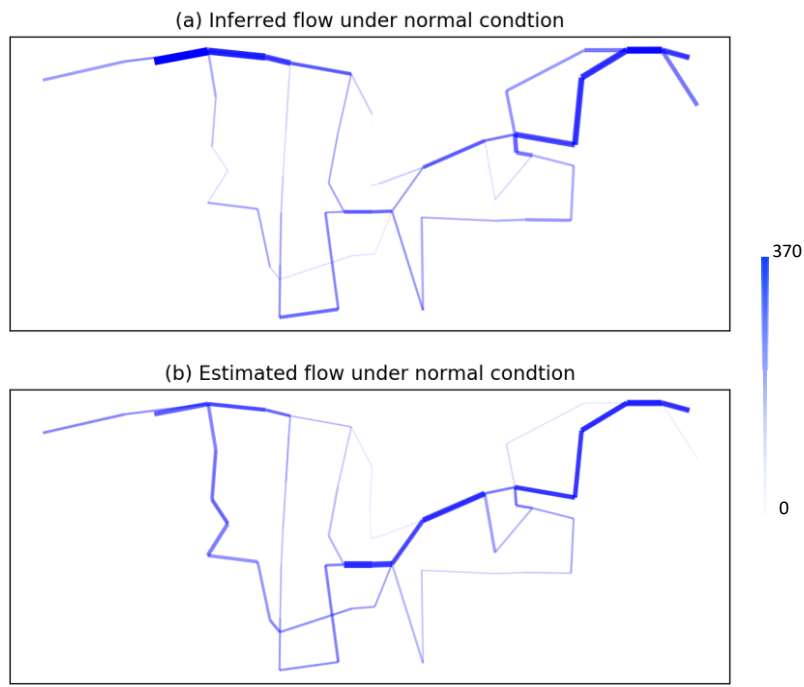

Figure 14. (a) Observed transit hourly flow inferred using individual level data; (b) hourly flow estimated by the model.

## 6 CONCLUSION

Transit flow estimation is important for transit planning and operations. In many transit systems, the only available information is the aggregated count data. This study focuses on the problem of dynamic transit flow estimation for schedule-based systems based on count data. Two key contributions are made.

First, we made modifications to the existing schedule-based assignment model to be more theoretically rigorous. Modifications on the definition of preference set and state, on the calculation of optimal value function, and on the decision of preference sets are proposed.

Second, we proposed two dynamic transit flow estimation methods (SDMSA and DSDMSA) based on the schedule-based user equilibrium concept. DSDMSA is a streamlined version of SDMSA and hence runs faster. The computational complexity is $O(N^2 L)$ which makes it possible to apply the model in an online setting. Detouring upstream phenomena may occur on schedule-based transit networks.

The models are first tested on a small network. DSDMSA is shown to be ineffective while SDMSA is effective. The estimation errors (0.076 for the minute-level OD demand, 156.7 for hourly OD demand and 11.6 for ridership) are satisfactory. The sensitivity of SDMSA to measurement period is also studied; the measurement period is recommended to be 15 minute or less. Testing results for model robustness with respect to user perception show that the estimation model is effective when the perception errors are within the interval (60%, 160%).

The estimation model is tested using a realistic network. The estimated transit segment flow has a mean of 94.9 when the observed flow has a mean of 92.5 (difference of 2.5%); the standard error of estimation is 33.1 and the relative error is 42.5%.

For future research, this methodology may be extended to centralized time-dependent flow estimation for Mobility-as-a-Service systems. The stop count data can be considered with multiple other data sources. Also, this research may be used to study mobility providers' decision to share information to get a better holistic picture. Lastly, real-time flow estimation methodology for non-schedule-based transit systems may be studied to complement this research.

## ACKNOWLEDGMENTS

Thanks are due to Haiyang Liu, Jian Wang and Chi Xie for sharing data with us. The authors are partially supported by the C2SMART University Transportation Center (USDOT # 69A3551747124) and NSF CMMI-1652735.

Trajectory with Smart Card Data and Train Schedules. *Sustainability,* 12**,** 2574.

ZHANG, J., SHEN, D., TU, L., ZHANG, F., XU, C., WANG, Y., TIAN, C., LI, X., HUANG, B. & LI, Z. 2017. A real-time passenger flow estimation and prediction method for urban bus transit systems. *IEEE Transactions on Intelligent Transportation Systems,* 18**,** 3168-3178.

ZHU, Y., KOUTSOPOULOS, H. N. & WILSON, N. H. 2017. A probabilistic Passenger-to-Train Assignment Model based on automated data. *Transportation Research Part B: Methodological,* 104**,** 522-542.